\documentclass{aastex61}

\received{}
\revised{}
\accepted{}
\submitjournal{ApJ}

\begin{document}

\title{Observations of cyanopolyynes toward four high-mass star-forming regions containing hot cores}

\correspondingauthor{Kotomi Taniguchi}
\email{kotomi.taniguchi@nao.ac.jp}

\author{Kotomi Taniguchi}
\altaffiliation{Research Fellow of Japan Society for the Promotion of Science}
\affiliation{Department of Astronomical Science, School of Physical Science, SOKENDAI (The Graduate University for Advanced Studies), Osawa, Mitaka, Tokyo 181-8588, Japan}
\affiliation{Nobeyama Radio Observatory, National Astronomical Observatory of Japan, Minamimaki, Minamisaku, Nagano 384-1305, Japan}

\author{Masao Saito}
\altaffiliation{The present address : National Astronomical Observatory of Japan, Osawa, Mitaka, Tokyo 181-8588, Japan}
\affiliation{Nobeyama Radio Observatory, National Astronomical Observatory of Japan, Minamimaki, Minamisaku, Nagano 384-1305, Japan}
\affiliation{Department of Astronomical Science, School of Physical Science, SOKENDAI (The Graduate University for Advanced Studies), Osawa, Mitaka, Tokyo 181-8588, Japan}

\author{Tomoya Hirota}
\affiliation{National Astronomical Observatory of Japan, Osawa, Mitaka, Tokyo 181-8588, Japan}
\affiliation{Department of Astronomical Science, School of Physical Science, SOKENDAI (The Graduate University for Advanced Studies), Osawa, Mitaka, Tokyo 181-8588, Japan}

\author{Hiroyuki Ozeki}
\affiliation{Department of Environmental Science, Faculty of Science, Toho University, Miyama, Funabashi, Chiba 274-8510, Japan}

\author{Yusuke Miyamoto}
\affiliation{Nobeyama Radio Observatory, National Astronomical Observatory of Japan, Minamimaki, Minamisaku, Nagano 384-1305, Japan}

\author{Hiroyuki Kaneko}
\affiliation{Nobeyama Radio Observatory, National Astronomical Observatory of Japan, Minamimaki, Minamisaku, Nagano 384-1305, Japan}

\author{Tetsuhiro Minamidani}
\affiliation{Nobeyama Radio Observatory, National Astronomical Observatory of Japan, Minamimaki, Minamisaku, Nagano 384-1305, Japan}
\affiliation{Department of Astronomical Science, School of Physical Science, SOKENDAI (The Graduate University for Advanced Studies), Osawa, Mitaka, Tokyo 181-8588, Japan}

\author{Tomomi Shimoikura}
\affiliation{Department of Astronomy and Earth Sciences, Tokyo Gakugei University, Nukuikitamachi, Koganei, Tokyo 184-8501, Japan}

\author{Fumitaka Nakamura}
\affiliation{National Astronomical Observatory of Japan, Osawa, Mitaka, Tokyo 181-8588, Japan}
\affiliation{Department of Astronomical Science, School of Physical Science, SOKENDAI (The Graduate University for Advanced Studies), Osawa, Mitaka, Tokyo 181-8588, Japan}

\author{Kazuhito Dobashi}
\affiliation{Department of Astronomy and Earth Sciences, Tokyo Gakugei University, Nukuikitamachi, Koganei, Tokyo 184-8501, Japan}



\begin{abstract}

We carried out line survey observations at the 26$-$30 GHz band toward the four high-mass star-forming regions containing hot cores, G10.30-0.15, G12.89+0.49, G16.86-2.16, and G28.28-0.36, with the Robert C. Byrd Green Bank Telescope.
We have detected HC$_{5}$N from all of the sources, and HC$_{7}$N from the three sources, except for G10.30-0.15.
We further conducted observations of HC$_{5}$N at the 42$-$46 GHz and 82$-$103 GHz bands toward the three sources, G12.89+0.49, G16.86-2.16, and G28.28-0.36, with the Nobeyama 45 m radio telescope.
The rotational lines of HC$_{5}$N with the high excitation energies ($E_{\rm {u}}/k \sim 63-100$ K), which are hardly excited in the cold dark clouds, have been detected from the three sources.
The rotational temperatures of HC$_{5}$N are found to be $\sim 13-20$ K in the three sources.
The detection of the lines with the high excitation energies and the derived rotational temperatures indicate that HC$_{5}$N exists in the warm gas within 0.07$-$0.1 pc radii around massive young stellar objects.
The column densities of HC$_{5}$N in the three sources are derived to be ($\sim 2.0-2.8$) $\times 10^{13}$ cm$^{-2}$.
We compare the ratios between $N$(HC$_{5}$N) the column density of HC$_{5}$N and $W$(CH$_{3}$OH) the integrated intensity of the thermal CH$_{3}$OH emission line among the three high-mass star-forming regions.
We found a possibility of the chemical differentiation in the three high-mass star-forming regions; G28.28-0.36 shows the largest $N$(HC$_{5}$N)/$W$(CH$_{3}$OH) ratio of $> 8.0 \times 10^{14}$ in units of (K km s$^{-1}$)$^{-1}$ cm$^{-2}$, while G12.89+0.49 and G16.86-2.16 show the smaller values ($\sim 2 \times 10^{13}$). 

\end{abstract}




\section{Introduction} \label{sec:intro}

As the star formation process progresses, the chemical composition of molecular gas changes. 
The star formation starts with the cold dark molecular clouds ($T \sim 10$ K, $n \sim 10^{4}$ cm$^{-3}$).
In the dark clouds, carbon-chain molecules such as CCS and HC$_{5}$N are abundant \citep{1992apj...392...551, 2009apj...699...585}.
After formation of protostars, the cloud temperature rises.
Such physical changes significantly affect the chemical composition, leading to hot core chemistry in the high-mass star-forming regions and hot corino chemistry in the low-mass star-forming regions.
In hot core and hot corino regions ($T \geq 100$ K, $n \sim 10^{6}$ cm$^{-3}$), while carbon-chain molecules except for HC$_{3}$N are deficient, saturated complex organic molecules (COMs) such as CH$_{3}$OH and CH$_{3}$CN are rich.
These COMs are formed by the combination of the grain-surface reactions and the following gas-phase reactions \citep[e.g.][]{2006A&A...457...927, 2009araa...47...427}.

Conversely, two low-mass star-forming cores with a variety of carbon-chain molecules have been discovered; L1527 in Taurus \citep{2008apj...672...371} and IRAS 15398-3359 in Lupus \citep{2009apj...697...769}.
The rotational temperatures of carbon-chain molecules in L1527 were derived to be $\sim12-15$ K \citep{2009apj...702...1025}, which are higher than those in the cold dark clouds.
\citet{2008ApJ...674...993} and \citet{2008ApJ...681...1385} showed that CH$_{4}$, which is evaporated from grain mantles in the warm regions ($T \sim 20$ K) around the protostar, could form carbon chains in the gas phase.
In fact, carbon-chain molecules were found to be enhanced in the warm regions where the temperature becomes higher than 20$-$30 K \citep{2010ApJ...722.1633}.
Such a chemical mechanism was named warm carbon chain chemistry \citep[WCCC,][]{2008apj...672...371}.

While there have been many studies about carbon-chain molecules in the low-mass star-forming regions, our knowledge of carbon-chain molecules in the high-mass star-forming regions is still poor.
As with the low-mass star-forming cores, understanding chemical mechanisms of carbon-chain molecules in the high-mass star-forming cores is one of the keys to revealing the initial conditions of massive star formation.
Cyanoacetylene (HC$_{3}$N), the shortest member of cyanopolyynes, is one of the hot core tracers \citep[a dense gas tracer; $n_{\rm {H}_{2}} \sim10^{5}-10^{6}$ cm$^{-3}$,][]{1996ApJ...460...343}, whereas longer cyanopolyynes are rarely detected.
\citet{2009MNRAS...394...221} demonstrated their chemical model calculation, and suggested that cyanopolyynes (HC$_{2n+1}$N; $n=1-4$) could be efficiently formed by the neutral-neutral reactions between CN and C$_{2n}$H$_{2}$.
The latter is formed from C$_{2}$H$_{2}$ evaporated from grain mantles in hot core regions. 
\citet{2016apj...830...106} proposed that the main formation pathway of HC$_{3}$N in the high-mass star-forming region G28.28-0.36 is the neutral-neutral reaction between C$_{2}$H$_{2}$ and CN, which is the same reaction suggested by \citet{2009MNRAS...394...221}, based on their observations of the $^{13}$C isotopic fractionation with the Nobeyama 45 m radio telescope.
Motivated by the chemical model calculation of \citet{2009MNRAS...394...221}, \citet{2014MNRAS...443...2252} conducted survey observations of HC$_{5}$N using the $J=12-11$ rotational line toward 79 hot cores with the Tidbinbilla 34 m radio telescope, and detected HC$_{5}$N toward 35 sources.
They could not derive the excitation temperatures due to the detection of the single rotational line which can be excited even in the cold dark clouds.
Thus, the detected emission lines may come from the cold envelopes around the protostars.
We still cannot confirm that long cyanopolyynes are formed and exist in hot cores.

In this paper, we carried out observations at the 26$-$30 GHz band toward four high-mass star-forming regions containing hot cores, G10.30-0.15, G12.89+0.49, G16.86-2.16, and G28.28-0.36, with the Robert C. Byrd Green Bank Telescope (GBT)\footnote{The National Radio Astronomy Observatory is a facility of the National Science Foundation operated under cooperative agreement by Associated Universities, Inc.}.
Two rotational transitions of HC$_{5}$N ($J = 10-9$ and 11$-$10) and three transitions of HC$_{7}$N ($J = 24-23$, 25$-$24, and  26$-$25) are in the frequency band.
HC$_{7}$N has been detected in several star-forming regions \citep{2008apj...672...371, 2012ApJ...744..131C, 2013MNRAS.436.1513F, 2015aap...581...A71}, which is considered as an indicator of cyanopolyyne-rich sources.
We also observed the rotational lines of HC$_{5}$N at the 42$-$46 GHz and 82$-$103 GHz bands toward the three sources, except for G10.30-0.15, with the Nobeyama 45 m radio telescope\footnote{The Nobeyama 45 m radio telescope is operated by the Nobeyama Radio Observatory, a branch of the National Astronomical Observatory of Japan.}.
Several lines of HC$_{5}$N with the high excitation energies ($E_{\rm {u}}/k \sim 63-100$ K) are in the 82$-$103 GHz band.
Such lines are hardly excited in the cold environments, and the detection of these lines means that HC$_{5}$N exists in the warm regions, as discussed later.
From the results with the two radio telescopes, we derive the rotational temperatures and the column densities of HC$_{5}$N in the three sources using the rotational diagram.
We confirm that HC$_{5}$N exists in the warm gas in the high-mass star-forming regions from the rotational temperatures.
The thermal emission line of CH$_{3}$OH, which is one of the hot core tracers and representative COMs, is in the frequency band of the GBT observations.
The origin of CH$_{3}$OH is considered to be different from that of carbon-chain species.
We then investigate the relationship between HC$_{5}$N and CH$_{3}$OH in the three high-mass star-forming regions in Section \ref{sec:diversity1}.
We found a possibility of the chemical differentiation among the three sources.

\section{Observations} \label{sec:obs}

We describe the observational details of each frequency band in this section.
Table \ref{tab:obs} summarizes the observing parameters of each frequency band.

The observed positions, distances, systemic velocities, and related objects (ultracompact \ion{H}{2} regions and outflows) of our four targets are summarized in Table \ref{tab:tab1}.
We selected these targets from the list of sources detected in HC$_{5}$N \citep{2014MNRAS...443...2252} by applying the following three criteria:
\begin{enumerate}
\item the source declination is above -21\arcdeg,
\item the distance ($D$) is within 3 kpc, and
\item CH$_{3}$CN was detected \citep{2006MNRAS...367...553}.
\end{enumerate}
The first criterion enables us to observe with the GBT and the Nobeyama 45 m telescope.
The second criterion is required for preventing from heavy beam dilution.
The last one ensures that target sources contain hot cores.
CH$_{3}$CN is one of the hot core tracers, and it is always detected in the hot gas \citep[$> 100$ K,][]{2007aap...465...913}.
Eight sources in the list meet the three criteria. 
Since our purpose is to derive the accurate rotational temperatures, we chose three sources among the eight selected sources which show the highest peak intensities of HC$_{5}$N (G12.89+0.49, G16.86-2.16, and G28.28-0.36).
In order to investigate different conditions, we selected one source which shows the low peak intensity of HC$_{5}$N (G10.30-0.15) with the high peak intensities of CH$_{3}$CN.

\floattable
\begin{deluxetable}{ccccccc}
\tablecaption{Observing parameters\label{tab:obs}}
\tablewidth{0pt}
\tablehead{
\colhead{Frequency} & \colhead{Telescope} & \colhead{Beam size} & \colhead{$\eta_{\rm {mb}}$} & \colhead{$T_{\rm {sys}}$} & \colhead{$\Delta \nu$} & \colhead{$T_{\rm {rms}}$} \\
\colhead{(GHz)} & \colhead{} & \colhead{($^{\prime\prime}$)} & \colhead{(\%)} & \colhead{(K)} & \colhead{(kHz)} & \colhead{(mK)}
}
\startdata
26$-$30 & GBT & 27 & 77 & 70$-$100 (April) & 66 & 10$-$17 ($T_{\rm {mb}}$) \\
 & & & & 90$-$120 (July) & & \\
42$-$46 & Nobeyama & 37 & 71 & 120$-$150 & 122.07 & 6$-$14 ($T_{\rm {A}}^{*}$) \\
82$-$103 & Nobeyama & 18\tablenotemark{a} & 54\tablenotemark{a} & 120$-$200 & 244.14 & 3$-$6 ($T_{\rm {A}}^{*}$) \\
\enddata
\tablenotetext{a}{The values are at the 86 GHz.}
\end{deluxetable}

\floattable
\begin{deluxetable}{ccccccc}
\tablecaption{Properties of our target sources\label{tab:tab1}}
\tablewidth{0pt}
\tablehead{
\colhead{Source} & \colhead{R.A.\tablenotemark{a}} & \colhead{Decl.\tablenotemark{a}} &\colhead{$D$} & \colhead{$V_{\rm {LSR}}$\tablenotemark{a}} & \multicolumn{2}{c}{Other Association\tablenotemark{b}} \\
\cline{6-7}
\colhead{} & \colhead{(J2000)} & \colhead{(J2000)} & \colhead{(kpc)} & \colhead{(km s$^{-1}$)} & \colhead{UCHII\tablenotemark{a}} & \colhead{outflow} 
}
\startdata
G10.30-0.15 & 18$^{\rm h}$08$^{\rm m}$55\fs5 & -20\arcdeg05\arcmin58\farcs0 & 2.1\tablenotemark{c} & 13.0 & Y & N\tablenotemark{f} \\
G12.89+0.49 & 18$^{\rm h}$11$^{\rm m}$51\fs4 & -17\arcdeg31\arcmin30\farcs0 & 2.50\tablenotemark{d} & 33.3 & N & Y\tablenotemark{e} \\
G16.86-2.16 & 18$^{\rm h}$29$^{\rm m}$24\fs4 & -15\arcdeg16\arcmin04\farcs0 & 1.67\tablenotemark{e} & 17.8 & N & Y\tablenotemark{e} \\
G28.28-0.36 & 18$^{\rm h}$44$^{\rm m}$13\fs3 & -04\arcdeg18\arcmin03\farcs0 & 3.0\tablenotemark{c} & 48.9 & Y & Y\tablenotemark{f} \\
\enddata
\tablenotetext{a}{\citet{2006MNRAS...367...553}}
\tablenotetext{b}{The symbols of "Y" and "N" represent detection and non-detection, respectively. ``UCHII" indicates an ultracompact \ion{H}{2} region lies within the Mopra beam ($\sim 38$$^{\prime\prime}$). The 6.7 GHz methanol masers are associated with all of the four sources.}
\tablenotetext{c}{\citet{2014MNRAS...443...2252}}
\tablenotetext{d}{\citet{2014apj...783...130}}
\tablenotetext{e}{\citet{2016aj...152...92L}}
\tablenotetext{f}{\citet{2008AJ...136...2391}}
\end{deluxetable}

\subsection{Observations with the Robert C. Byrd Green Bank Telescope (GBT)} \label{sec:obsGBT}

We carried out observations with the Robert C. Byrd Green Bank Telescope (GBT) of the National Radio Astronomy Observatory in 2016 April and July.
We used the Ka-band receiver to cover the 26.1$-$29.7 GHz range, except for 28.7$-$28.9 GHz.
The beam size (HPBW), the aperture efficiency ($\eta_{\rm {A}}$), and the main beam efficiency ($\eta_{\rm {mb}}$) of the Ka-band receiver were approximately 27$^{\prime\prime}$, 67\%, and 77\%, respectively.
The typical system temperatures were from 70 to 100 K in April, and from 90 to 120 K in July, depending on the weather conditions and elevations.  
We used the VEGAS spectrometer whose bandwidth and frequency resolution were 1080 MHz and 66 kHz, respectively.
The frequency resolution of 66 kHz corresponds to the velocity resolution of 0.73 km s$^{-1}$ at 27 GHz.
The center frequencies of four spectral windows were set at 26.525, 27.425, 28.325, and 29.275 GHz, respectively.

We employed the position-switching mode.
The integration time was 60 seconds per on-source and off-source positions.
The observed positions of the four target sources are summarized in Table \ref{tab:tab1}.
The off-source positions were set to be +15$^{\prime}$ away in the declination. 

The pointing sources were J1850-0001 at ($\alpha_{2000}$, $\delta_{2000}$) = (18$^{\rm h}$50$^{\rm m}$31\fs17, -00\arcdeg01\arcmin55\farcs1) for G28.28-0.36, and J1833-2103 at ($\alpha_{2000}$, $\delta_{2000}$) = (18$^{\rm h}$33$^{\rm m}$39\fs93, -21\arcdeg03\arcmin40\farcs769) for the others.
We checked the pointing accuracy every 1.5-2 hours, and the pointing error was less than 8$^{\prime\prime}$, depending on the weather conditions.
We observed the bright radio continuum source 3C 286 at ($\alpha_{2000}$, $\delta_{2000}$) = (13$^{\rm h}$31$^{\rm m}$08\fs288, +30\arcdeg30\arcmin32\farcs96) in the beginning of every observing session in order to conduct the absolute flux calibration. 
The intensity calibration error was less than 5\%.

\subsection{Observations with the Nobeyama 45 m radio telescope} \label{sec:obsNRO}

The observations were carried out between 2017 January and March.
We observed toward the three sources G12.89+0.49, G16.86-2.16, and G28.28-0.36.
The on-source and off-source positions were set at the same ones with the GBT observations.
The position-switching mode was employed, and the integration time was 20 seconds per scan.

We used the Z45 receiver \citep{2015PASJ...67..117N} for the 42$-$46 GHz band observations.
The system temperatures of the Z45 receiver were 120$-$150 K, depending on the weather conditions and elevations.
The main beam efficiency ($\eta_{\rm {mb}}$) and the beam size (HPBW) at 43GHz were 71\% and 37$^{\prime\prime}$, respectively.
The TZ receiver \citep{2013PASP..125..252N} was used for the 82$-$103 GHz band observations.
The system temperatures were from 120 to 200 K.
The main beam efficiency and the beam size at 86 GHz were 54\% and 18$^{\prime\prime}$, respectively.

We used the SAM45 FX-type digital correlator \citep{2012PASJ...64...29K} in frequency settings whose bandwidths and resolutions are 500 MHz and 122.07 kHz for observations of the Z45 receiver, and 1000 MHz and 244.14 kHz for those of the TZ receiver, respectively.
The frequency resolutions correspond to the velocity resolution of $\sim 0.85$ km s$^{-1}$.

We checked the telescope pointing every 1.5 hr by observing SiO maser line ($J=1-0$) from OH39.7+1.5 at ($\alpha_{2000}$, $\delta_{2000}$) = (18$^{\rm h}$56$^{\rm m}$03\fs88, +06\arcdeg38\arcmin49\farcs8).
We used the Z45 receiver for the 42$-$46 GHz band observations, and the H40 receiver for the 82$-$103 GHz band observations. 
The pointing error was less than 3$^{\prime\prime}$.
\\
\section{Results and Analyses}

\subsection{Results}

\subsubsection{Observational results with the GBT}
\begin{figure}
\figurenum{1}
\plotone{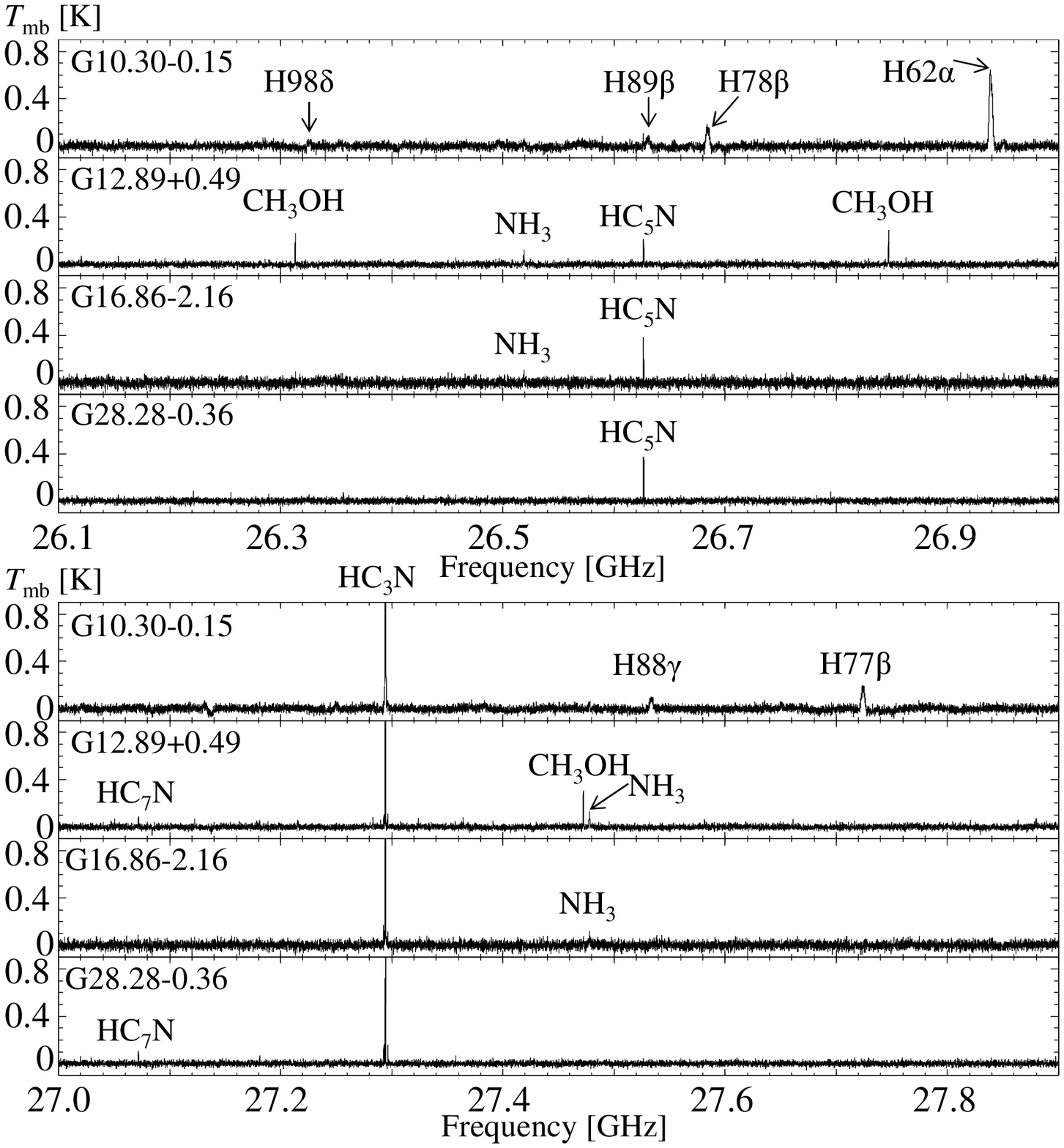}
\caption{Spectra toward the four hot cores from 26.1 to 27.0 GHz (upper) and from 27.0 to 27.9 GHz (lower) with the GBT. The source names are presented at upper left parts of each panel. The frequency resolution is 66 kHz.\label{fig:f1}}
\end{figure}

\begin{figure}
\figurenum{2}
\plotone{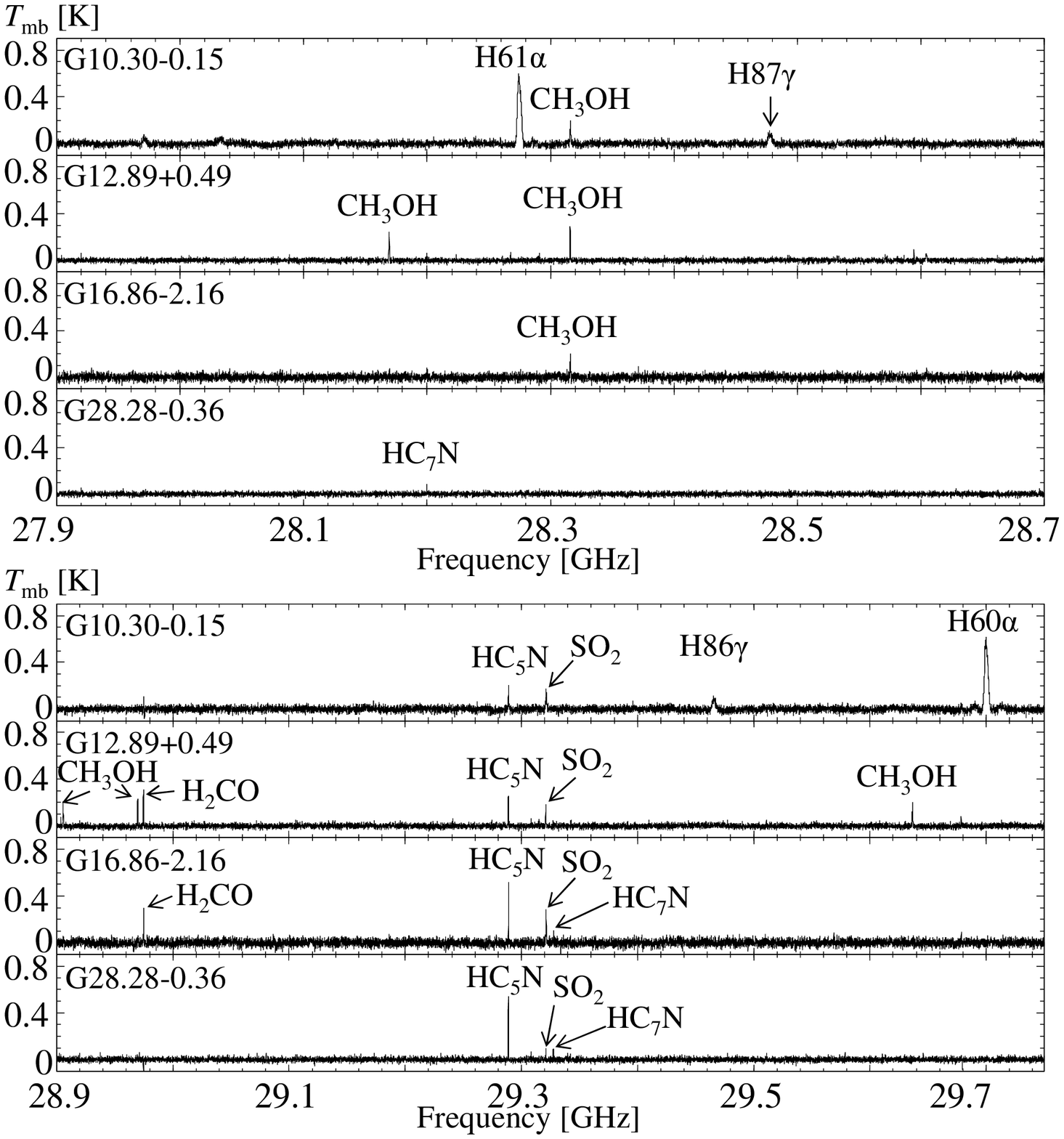}
\caption{Spectra toward the four hot cores from 27.9 to 28.7 GHz (upper) and from 28.9 to 29.7 GHz (lower) with the GBT. The source names are presented at upper left parts of each panel. The frequency resolution is 66 kHz.\label{fig:f2}}
\end{figure}

We conducted data reduction\footnote{We set the velocity units as ``lsrk'' in the observational scripts, but we wrote the ``$V_{\rm {LSR}}$'' values in the source catalog.
We then shifted the frequency so that the emission peaks of HC$_{3}$N correspond to its rest frequency, when we made the final spectra.} using the GBTIDL\footnote{http://gbtidl.nrao.edu}.
The overviews of obtained spectra toward the four sources are shown in Figures \ref{fig:f1} and \ref{fig:f2}.
The rms noise levels are 14, 10, 17, and 11 mK in $T_{\rm {mb}}$ for G10.30-0.15, G12.89+0.49, G16.86-2.16, and G28.28-0.36, respectively.
We fitted the spectra with a Gaussian profile and obtained the spectral line parameters, as summarized in Table \ref{tab:tab2}.
The lower detection limit is set to be $4 \sigma$.

Two and three rotational lines of HC$_{5}$N  ($J=10-9$ and $11-10$) and HC$_{7}$N ($J=24-23$, $25-24$, and $26-25$) are in the observed frequency range, respectively.
All of the observed rotational lines of HC$_{5}$N were detected from all of the four sources.
We also detected HC$_{7}$N from the three sources, except for G10.30-0.15.
Detection of HC$_{7}$N has been reported in several star-forming regions \citep{2008apj...672...371, 2012ApJ...744..131C, 2013MNRAS.436.1513F, 2015aap...581...A71}. 
The detection of HC$_{7}$N from the three high-mass star-forming regions suggests that our target cores contain the long cyanopolyynes abundantly.

In addition, several other lines were detected. 
The hydrogen recombination lines are prominent only in G10.30-0.15, which seems to be caused by a few ultracompact \ion{H}{2} (UC\ion{H}{2}) regions associated with G10.30-0.15 \citep{2006aap...453...1003}.
Although a UC\ion{H}{2} region is associated with G28.28-0.36 \citep{2003aap...410...597}, we did not detect any hydrogen recombination lines in G28.28-0.36.
The rotational lines of HC$_{3}$N were detected toward all of the four sources.
HC$_{3}$N is considered as one of the hot core tracers (a dense gas tracer), and it is usually detected in hot core regions, as mentioned in Section \ref{sec:intro}. 
The thermal CH$_{3}$OH emission lines ($J_{K}=4_{0}-3_{1}$ $E$) with the low-excitation energy ($E_{\rm {u}}/k = 36.3$ K) were detected except for G28.28-0.36, and its lines with high-excitation energies ($E_{\rm {u}}/k \sim 120-340$ K) were detected only toward G12.89+0.49.
Metastable inversion transition lines of NH$_{3}$ with very high-excitation energies, i.e., ($J$, $K$) = (8, 8) at 26.51898 GHz ($E_{\rm {u}}/k \sim 686$ K) and ($J$, $K$) = (9, 9) at 27.47794 GHz ($E_{\rm {u}}/k \sim 852$ K), were detected toward G12.89+0.49 and G16.86-2.16.
The results indicate that extremely hot gas is contained in these regions \citep{2011apj...739...L13}.
The H$_{2}$CO ($J_{Ka,Kc}=3_{1,2}-3_{1,3}$) emission lines were detected toward G12.89+0.49 and G16.86-2.16.

\subsubsection{Observational results with the Nobeyama 45 m radio telescope}

\begin{figure}
\figurenum{3}
\plotone{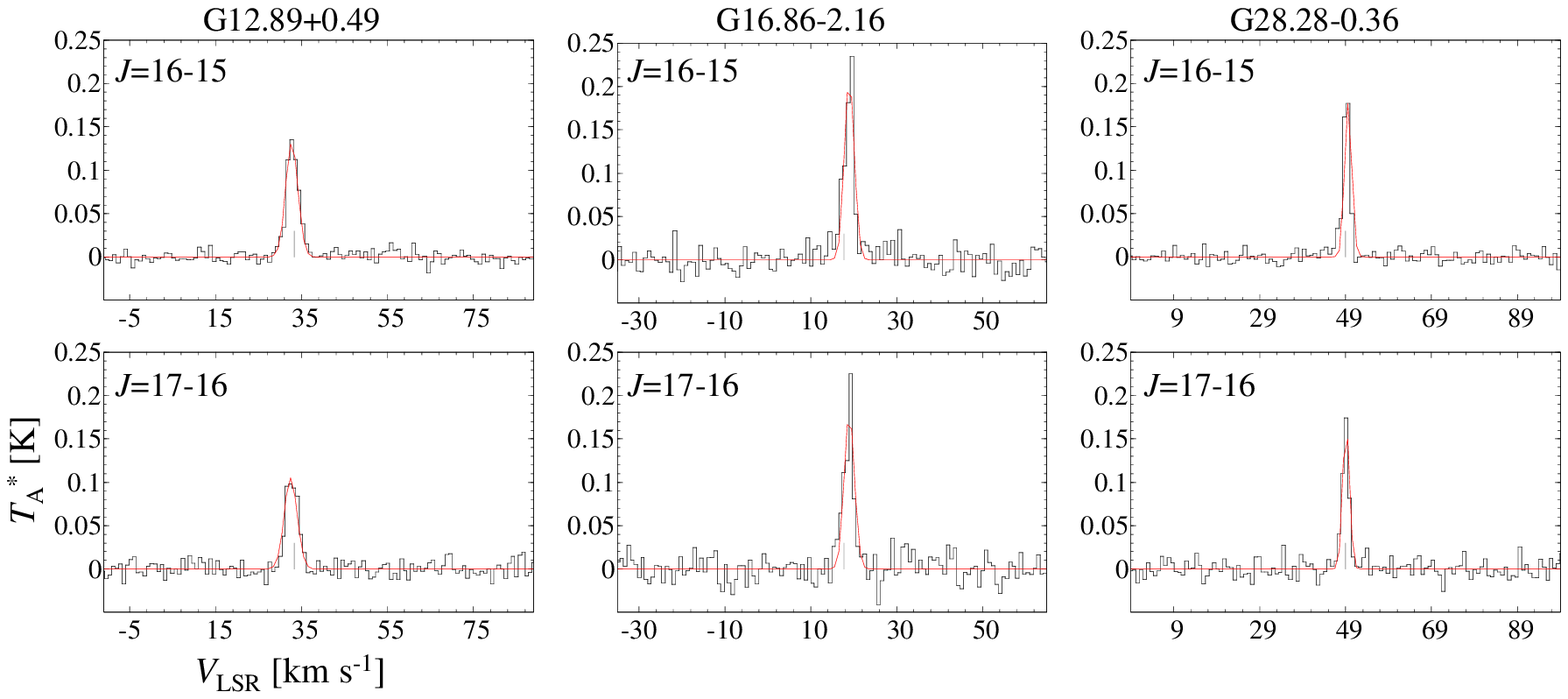}
\caption{Spectra of HC$_{5}$N at the 42$-$46 GHz band toward the three hot cores with the Nobeyama 45 m telescope. The red lines show the results of the Gaussian fitting, and the vertical lines indicate the systemic velocity of each source. \label{fig:f3}}
\end{figure}

\begin{figure}
\figurenum{4}
\plotone{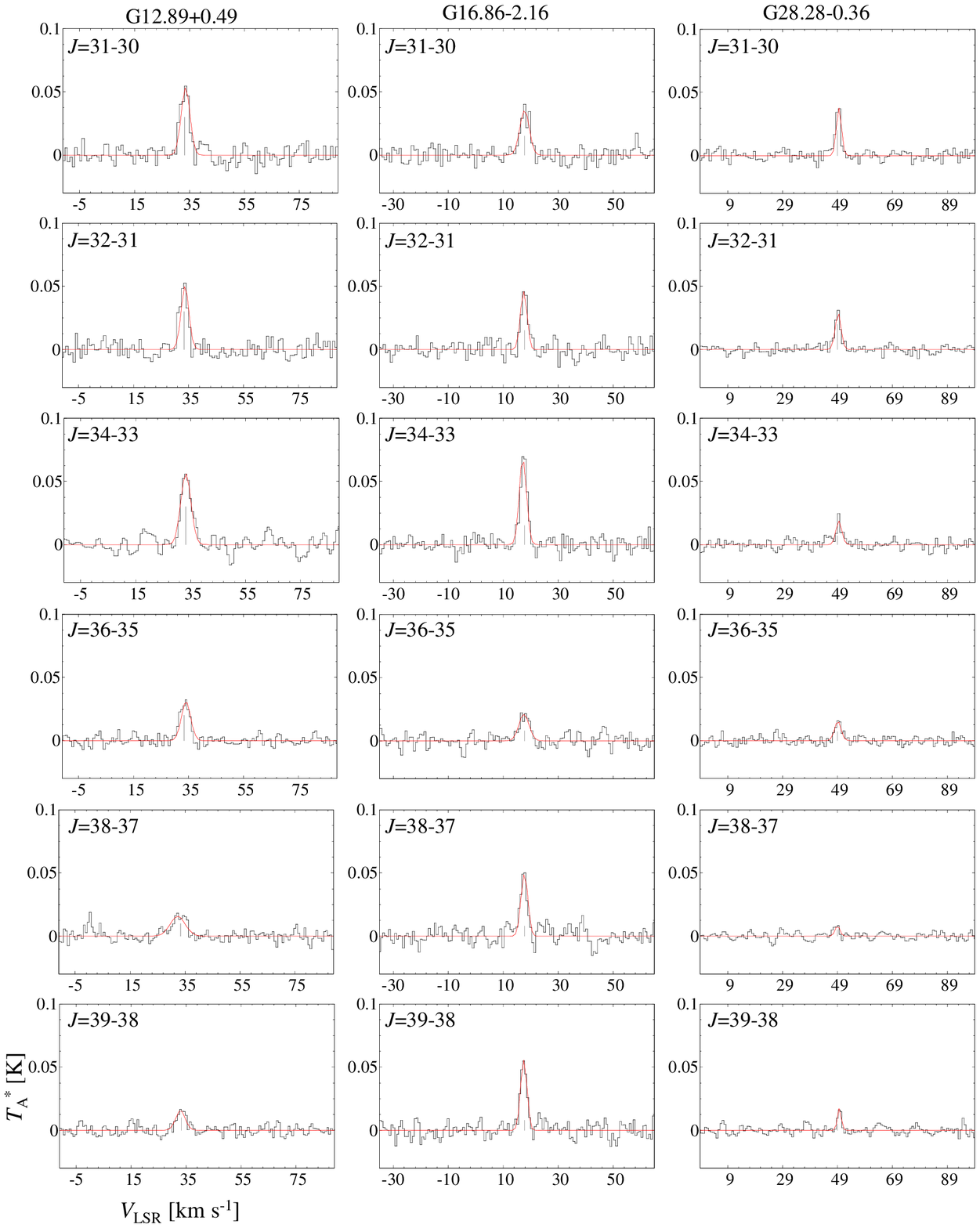}
\caption{Spectra of HC$_{5}$N at the 82$-$103 GHz band toward the three hot cores with the Nobeyama 45 m telescope. The red lines show the results of the Gaussian fitting, and the vertical lines indicate the systemic velocity of each source. \label{fig:f4}}
\end{figure}

We conducted data reduction using the Java Newstar, which is an astronomical data  analyzing system of Nobeyama data\footnote{http://www.nro.nao.ac.jp/~jnewstar/html/}.
Figures \ref{fig:f3} and \ref{fig:f4} show the spectra of HC$_{5}$N at the 42$-$46 GHz and 82$-$103 GHz bands, respectively.
The rms noise levels are from 6 to 14 mK in $T_{\rm {A}}^{*}$ at the 42$-$46 GHz band and from 3 to 6 mK at the 82$-$103 GHz band.
Table \ref{tab:tab3} summarizes the line parameters obtained from the Gaussian fitting.
The main beam efficiencies which we used in the analysis are summarized in Table \ref{tab:tab3}.
 
The signal-to-noise ratios were between 4 and 23.
The two rotational lines of HC$_{5}$N at the 42$-$46 GHz band ($J=16-15$ and 17$-$16) were detected from all of the three sources.
Moreover, we detected six rotational lines of HC$_{5}$N with the high excitation energies ($E_{\rm {u}}/k \sim 63-100$ K) from all of the sources.
The $V_{\rm {LSR}}$ values of the detected lines are summarized in Table \ref{tab:tab3}.
The $V_{\rm {LSR}}$ values of the HC$_{5}$N lines are consistent with the systemic velocities of each source (Table \ref{tab:tab1}).

\floattable
\rotate
\begin{deluxetable}{cccccccccccccccccc}
\tabletypesize{\scriptsize}
\tablecaption{Spectral line parameters observed with the GBT \label{tab:tab2}}
\tablewidth{0pt}
\tablehead{
\colhead{} & \colhead{} & \colhead{}  & \multicolumn{3}{c}{G10.30-0.15} & \colhead{} & \multicolumn{3}{c}{G12.89+0.49} & \colhead{} & \multicolumn{3}{c}{G16.86-2.16} & \colhead{} & \multicolumn{3}{c}{G28.28-0.36} \\
\cline{4-6}\cline{8-10}\cline{12-14}\cline{16-18}
\colhead{Species} & \colhead{Frequency\tablenotemark{a}} & \colhead{$E_{\rm {u}}/k$} & \colhead{$T_{\rm {mb}}$} & \colhead{FWHM} & \colhead{$\int T_{\mathrm {mb}}dv$} & \colhead{} & \colhead{$T_{\rm {mb}}$} & \colhead{FWHM} & \colhead{$\int T_{\mathrm {mb}}dv$} & \colhead{} & \colhead{$T_{\rm {mb}}$} & \colhead{FWHM} & \colhead{$\int T_{\mathrm {mb}}dv$} & \colhead{} & \colhead{$T_{\rm {mb}}$} & \colhead{FWHM} & \colhead{$\int T_{\mathrm {mb}}dv$} 
\\
\colhead{(Transition)} & \colhead{(GHz)} & \colhead{ (K)} & \colhead{(K)} & \colhead{(km s$^{-1}$)} & \colhead{(K km s$^{-1}$)} & \colhead{} & \colhead{(K)} & \colhead{(km s$^{-1}$)} & \colhead{(K km s$^{-1}$)} & \colhead{} & \colhead{(K)} & \colhead{(km s$^{-1}$)} & \colhead{(K km s$^{-1}$)} & \colhead{} & \colhead{(K)} & \colhead{(km s$^{-1}$)} & \colhead{(K km s$^{-1}$)}
}
\startdata
HC$_{3}$N & & & & & & & & & & & & & & & & & \\
($J=3-2$) & 27.29429 & 2.6 & 1.14 (5) & 7.9 (4) & 9.6 (6) & & 1.60 (11) & 4.5 (4) & 7.7 (8) & & 1.94 (5) & 4.42 (14) & 9.1 (4) & & 2.24 (11) & 2.38 (14) & 5.7 (4) \\
HC$_{5}$N & & & & & & & & & & & & & & & & & \\
($J=10-9$) & 26.62654 & 7.0 & 0.107 (14) & 2.9 (4) & 0.33 (7) & & 0.209 (13) & 3.5 (3) & 0.78 (8) & & 0.37 (2) & 2.6 (2) & 1.03 (11) & & 0.37 (3) & 2.08 (17) & 0.81 (9) \\
($J=11-10$) & 29.28915 & 8.4 & 0.158 (11) & 4.7 (4) & 0.79 (8) & & 0.283 (16) & 3.6 (2) & 1.10 (9) & & 0.42 (3) & 3.1 (3) & 1.41 (16) & & 0.53 (2) & 1.91 (9) & 1.08 (7) \\
HC$_{7}$N & & & & & & & & & & & & & & & & & \\
($J=24-23$) & 27.07181 & 16.2 & $< 0.042$ & ... & ... & & 0.076 (8) & 4.5 (5) & 0.36 (6) & & $< 0.051$ & ... & ... & & 0.086 (10) & 2.4 (3) & 0.22 (4) \\
($J=25-24$) & 28.19981 & 17.6 & $< 0.042$ & ... & ... & & 0.051 (8) & 3.1 (6) & 0.16 (4) & &$< 0.051$  & ... & ... & & 0.083 (12) & 2.0 (3) & 0.17 (4) \\
($J=26-25$) & 29.32777 & 19.0 & $< 0.042$ & ... & ... & & $< 0.030$ & ... & ... & & 0.104 (12) & 1.17 (15) & 0.13 (2) & & 0.098 (12) & 1.8 (2) & 0.19 (3) \\
 & & & & & & & & & & & & & & & & & \\
CH$_{3}$OH & & & & & & & & & & & & & & & & & \\
($J_{K}=4_{0}-3_{1}$ $E$) & 28.31607 & 36.3 & 0.186 (5) & 7.5 (4) & 1.48 (10) & & 0.259 (14) & 4.5 (3) & 1.25 (10) & & 0.180 (14) & 5.5 (5) & 1.06 (13) & & $< 0.033$ & ... & ...  \\
SO$_{2}$ & & & & & & & & & & & & & & & & & \\
($J_{Ka,Kc}=4_{0,4}-3_{1,3}$)  & 29.32133 & 9.2 & - & - & - & & 0.16 (4) & 5.1 (1.3) & 0.9 (3) & & 0.27 (3) & 4.8 (7) & 1.4 (2) & & 0.084 (12) & 2.6 (4) & 0.23 (5) \\
H$_{2}$CO & & & & & & & & & & & & & & & & & \\
($J_{Ka,Kc}=3_{1,2}-3_{1,3}$) & 28.97480 & 33.4 & $< 0.042$ & ... & ... & & 0.284 (14) & 4.5 (3) & 1.37 (10) & & 0.285 (19) & 3.5 (3) & 1.05 (10) & & $< 0.033$ & ...& ... \\
\enddata
\tablecomments{Numbers in the parentheses are the standard deviation of the Gaussian fit, expressed in units of the last significant digits. For example, 1.14 (5) means $1.14 \pm 0.05$. We cannot fit the line of SO$_{2}$ line for G10.30-0.15 well, as denoted by `-' marks. The upper limits correspond to the $3 \sigma$ limits.}
\tablenotetext{a}{Taken from the Cologne Database for Molecular Spectroscopy (CDMS) \citep{2005JMoSt...742...215}.}
\end{deluxetable}
\floattable
\rotate
\begin{deluxetable}{cccccccccccccccccc}
\tabletypesize{\scriptsize}
\tablecaption{Spectral line parameters of HC$_{5}$N observed with the Nobeyama 45 m telescope \label{tab:tab3}}
\tablewidth{0pt}
\tablehead{
\colhead{} & \colhead{} & \colhead{}  & \multicolumn{4}{c}{G12.89+0.49} & \colhead{} & \multicolumn{4}{c}{G16.86-2.16} & \colhead{} & \multicolumn{4}{c}{G28.28-0.36} & \colhead{} \\
\cline{4-7}\cline{9-12}\cline{14-17}
\colhead{Transition} & \colhead{Frequency\tablenotemark{a}} & \colhead{$E_{\rm {u}}/k$} & \colhead{$T_{\rm {mb}}$} & \colhead{FWHM} & \colhead{$\int T_{\mathrm {mb}}dv$} & \colhead{$V_{\rm{LSR}}$\tablenotemark{b}} & \colhead{} & \colhead{$T_{\rm {mb}}$} & \colhead{FWHM} & \colhead{$\int T_{\mathrm {mb}}dv$} &  \colhead{$V_{\rm{LSR}}$\tablenotemark{b}} & \colhead{} & \colhead{$T_{\rm {mb}}$} & \colhead{FWHM} & \colhead{$\int T_{\mathrm {mb}}dv$} &  \colhead{$V_{\rm{LSR}}$\tablenotemark{b}} & \colhead{$\eta_{\rm {B}}$} \\
\colhead{($J'-J"$)} & \colhead{(GHz)} & \colhead{(K)} & \colhead{(K)} & \colhead{(km s$^{-1}$)} & \colhead{(K km s$^{-1}$)} & \colhead{(km s$^{-1}$)} & \colhead{} & \colhead{(K)} & \colhead{(km s$^{-1}$)} & \colhead{(K km s$^{-1}$)} & \colhead{(km s$^{-1}$)} & \colhead{} & \colhead{(K)} & \colhead{(km s$^{-1}$)} & \colhead{(K km s$^{-1}$)} & \colhead{(km s$^{-1}$)} & \colhead{(\%)}
}
\startdata
16$-$15 & 42.60215 & 17.4 & 0.186 (7) & 3.36 (14) & 0.67 (4) & 32.7 & & 0.296 (17) & 2.69 (18) & 0.85 (7) & 19.0 & & 0.251 (14) & 1.97 (12) & 0.53 (4) & 49.6 & 71 \\
17$-$16 & 45.26472 & 19.6 & 0.148 (8) & 3.5 (2) & 0.56 (5) & 32.4 & & 0.253 (17) & 2.8 (2) & 0.75 (8) & 19.0 & & 0.239 (13) & 1.88 (11) & 0.48 (4) & 49.1 & 71 \\
31$-$30 & 82.53904 & 63.4 & 0.096 (7) & 3.8 (3) & 0.39 (4) & 33.6 & & 0.063 (6) & 4.0 (5) & 0.27 (4) & 17.8 & & 0.068 (6) & 2.2 (2) & 0.16 (2) & 49.5 & 56 \\
32$-$31 & 85.20134 & 67.5 & 0.092 (8) & 3.4 (3) & 0.34 (4) & 33.7 & & 0.084 (8) & 2.8 (3) & 0.25 (3) & 17.4 & & 0.053 (4) & 2.6 (2) & 0.144 (17) & 49.3 & 54 \\
34$-$33 & 90.52589 & 76.0 & 0.109 (7) & 4.4 (3) & 0.50 (5) & 33.3 & & 0.133 (7) & 3.1 (2) & 0.44 (4) & 17.1 & & 0.036 (5) & 2.5 (4) & 0.10 (2) & 49.5 & 52 \\
36$-$35 & 95.85034 & 85.1 & 0.063 (5) & 4.0 (3) & 0.27 (3) & 34.1 & & 0.042 (6) & 4.4 (7) & 0.20 (4) & 17.5 & & 0.031 (4) & 2.8 (5) & 0.09 (2) & 49.1 & 49 \\
38$-$37 & 101.17468 & 94.7 & 0.034 (4) & 6.5 (9) & 0.23 (4) & 32.3 & & 0.106 (9) & 2.9 (3) & 0.33 (4) & 17.7 & & 0.018 (5) & 2.1 (6) & 0.039 (16) & 49.0 & 46 \\
39$-$38 & 103.83682 & 99.7 & 0.035 (5) & 4.1 (6) & 0.15 (3) & 33.0 & & 0.124 (9) & 2.7 (2) & 0.35 (4) & 17.5 & & 0.039 (6) & 1.6 (3) & 0.069 (17) & 49.7 & 45\\
\enddata
\tablecomments{Numbers in the parentheses are the standard deviation of the Gaussian fit, expressed in units of the last significant digits.}
\tablenotetext{a}{Taken from the Cologne Database for Molecular Spectroscopy (CDMS) \citep{2005JMoSt...742...215}.}
\tablenotetext{b}{The errors were 0.85 km s$^{-1}$, which corresponds to the velocity resolution (Section \ref{sec:obsNRO}).}
\end{deluxetable}

\subsection{Analyses} \label{sec:ana}

\begin{figure}
\figurenum{5}
\plotone{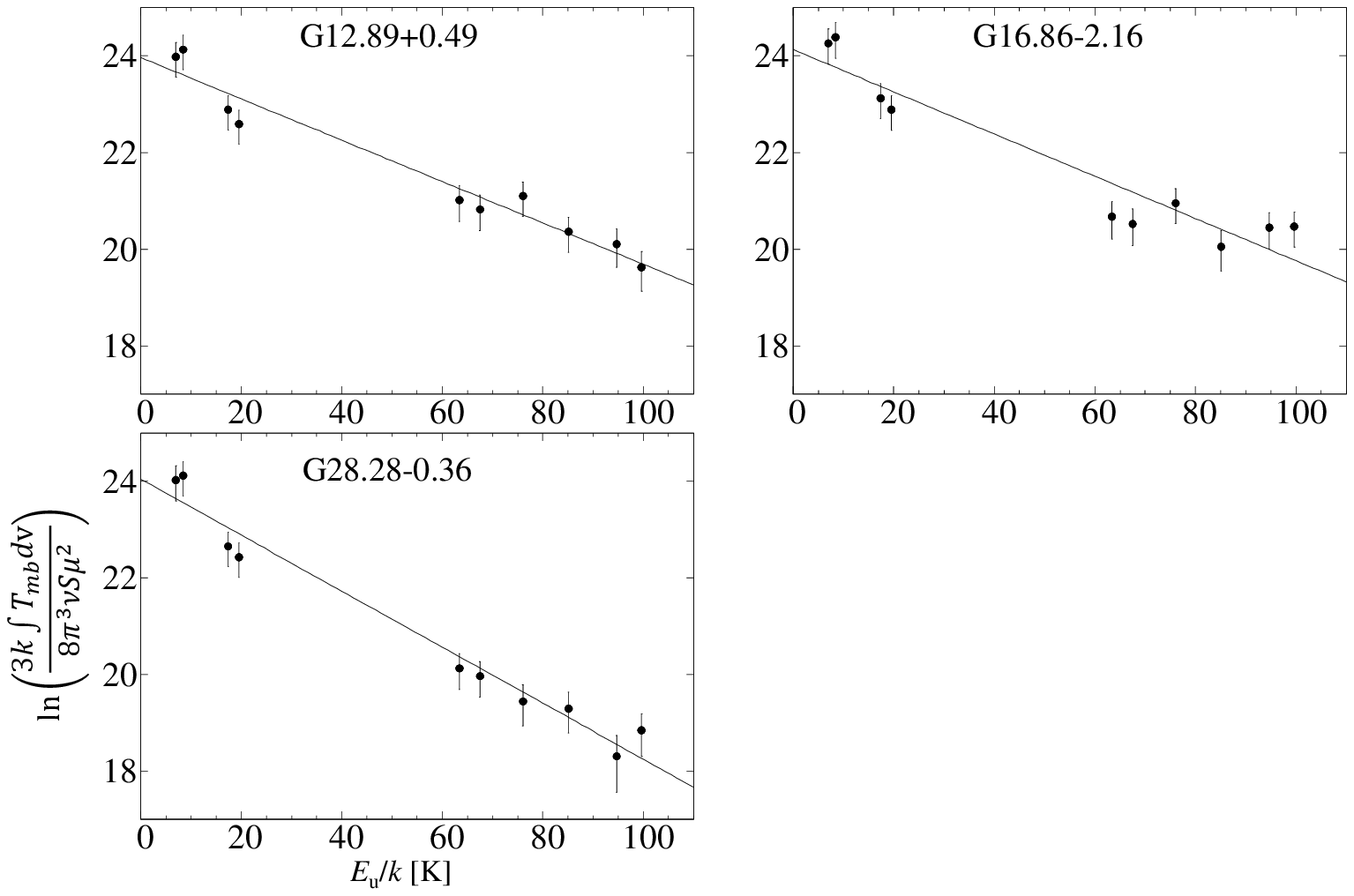}
\caption{Rotational diagram of HC$_{5}$N toward the three sources with the filling factor of unity. The error bars show one standard deviation. \label{fig:f5}}
\end{figure}

We derived the rotational temperatures and column densities of HC$_{5}$N in the three sources (G12.89+0.49, G16.86-2.16, and G28.28-0.36) from the rotational diagram analysis, using the following formula \citep{1999ApJ...517..209G};
\begin{equation} \label{rd}
{\rm {ln}} \frac{3k \int T_{\mathrm {mb}}dv}{8\pi ^3 \nu S \mu ^2} = {\rm {ln}} \frac{N}{Q(T_{\rm {rot}})} - \frac{E_{\rm {u}}}{kT_{\rm {rot}}},
\end{equation}
where $k$ is the Boltzmann constant, $S$ is the line strength, $\mu$ is the permanent electric dipole moment, $N$ is the column density, and $Q(T_{\rm {rot}})$ is the partition function.
We used 4.330 D \citep{1976JMoSp...62...175} for $\mu$ of HC$_{5}$N.
Figure \ref{fig:f5} shows the fitting results of the rotational diagram for the three sources.
The errors include the Gaussian fitting errors, the uncertainties from the main beam efficiency of 10\%, the chopper-wheel method of 10\%, and from the other factors such as calibration and filling factor of 30\%.
Since we do not know the spatial distributions of HC$_{5}$N, we assume the filling factor of unity.
In the fitting, the values of the $J=16-15$ and 17$-$16 transitions are systematically lower than the fitting lines.
This suggests a smaller filling factor than unity, because the beam size at the 42$-$46 GHz band with the Nobeyama 45 m telescope is significantly larger than those at the 26$-$30 GHz band with the GBT and at the 82$-$103 GHz band with the Nobeyama 45 m telescope.
The ranges of the derived rotational temperatures and column densities in the three sources are 17$-$23 K and (7.5$-$11)$\times 10^{12}$ cm$^{-2}$, as summarized in Table \ref{tab:tab4}.
In the case of G10.30-0.15, we detected only two lines with the similar energies using the GBT, and we cannot conduct the rotational diagram analysis.
We derived the column density of HC$_{5}$N in G10.30-0.15 to be $8.5 ^{+2.0}_{-1.7} \times 10^{12}$ cm$^{-2}$, using the average value of the rotational temperatures in the three sources ($21 \pm 11$ K).
The error of the rotational temperature in G10.30-0.15 was estimated from the relative errors of the rotational temperatures in the three sources, and the error of the column density corresponds to the change in the rotational temperature.

\begin{figure}
\figurenum{6}
\plotone{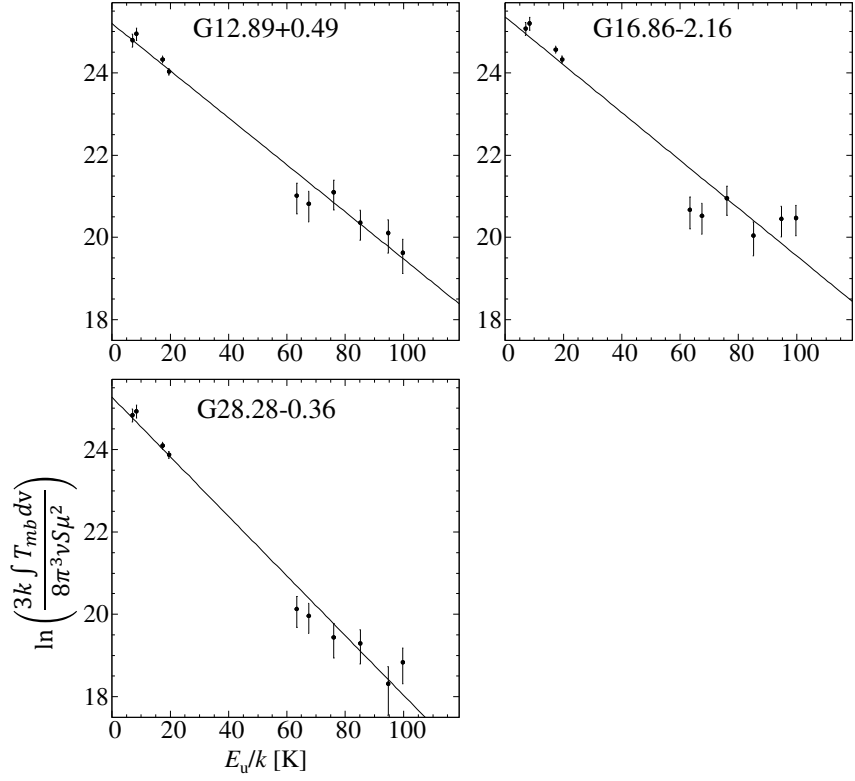}
\caption{Rotational diagram of HC$_{5}$N toward the three sources with the beam-size correction. The error bars show one standard deviation. \label{fig:f6}}
\end{figure}

Assuming a small beam filling factor, we multiplied the integrated intensities of the GBT data by ($\frac{27^{\prime\prime}}{18^{\prime\prime}}$)$^{2}$ and the 45GHz band data by ($\frac{37^{\prime\prime}}{18^{\prime\prime}}$)$^{2}$ for the correction of the different beam sizes.
Figure \ref{fig:f6} shows the rotational diagram with the beam-size correction.
All of the data, including the 42$-$46 GHz band, are better fitted.
From the above considerations of beam sizes and a small beam filling factor, the emission region sizes are limited to below 18$^{\prime\prime}$, which correspond to  0.07$-$0.1 pc radii at the target source distances (1.67$-$3.0 kpc, Table \ref{tab:tab1}).

Table \ref{tab:tab4} summarizes the derived values of the rotational temperatures and column densities.
The rotational temperatures are derived to be 13.8$-$18 K, and the corresponding column densities are (2.05$-$2.78) $\times 10^{13}$ cm$^{-2}$.
The derived rotational temperatures with and without the beam-size correction are consistent within their $1\sigma$ errors.
The rotational temperature and the column density in G10.30-0.15 were derived by the same manner as mentioned in the previous paragraph ($T_{\rm {rot}} = 16 \pm 4$ K, $N = 7.5 ^{+0.8}_{-0.6} \times 10^{12}$ cm$^{-2}$).
We use the results with the beam-size correction in the following sections, because the fitting results are better.

\floattable
\begin{deluxetable}{lcccc}
\tabletypesize{\scriptsize}
\tablecaption{The rotational temperatures and column densities of HC$_{5}$N in the four sources\label{tab:tab4}}
\tablewidth{0pt}
\tablehead{
\colhead{Source} & \multicolumn{2}{c}{filling factor = 1} & \multicolumn{2}{c}{With beam-size correction} \\
\cline{2-5} 
\colhead{} & \colhead{$T_{\rm {rot}}$ (K)} & \colhead{$N$ ($\times 10^{12}$ cm$^{-2}$)} & \colhead{$T_{\rm {rot}}$ (K)} & \colhead{$N$ ($\times 10^{12}$ cm$^{-2}$)}
}
\startdata
G10.30-0.15 & $21 \pm 11$\tablenotemark{a} & $8.5 ^{+2.0}_{-1.7}$ & $16 \pm 4$\tablenotemark{a} & $7.5 ^{+0.8}_{-0.6}$ \\
G12.89+0.49 & $23^{+7}_{-5}$ & $9.4 ^{+1.8}_{-2.0}$ & $18 \pm 2$ & $23.9 ^{+1.5}_{-1.7}$ \\
G16.86-2.16 & $23^{+7}_{-4}$ & $11 \pm 2 $ & $17 \pm 2$ & $27.8 ^{+1.6}_{-2.0}$ \\
G28.28-0.36 & $17^{+4}_{-3}$ & $7.5^{+1.9}_{-2.1}$ & $13.8 ^{+1.5}_{-1.1}$ & $20.5 ^{+2.0}_{-0.5}$ \\
\enddata
\tablecomments{The errors represent one standard deviation.}
\tablenotetext{a}{The value is the average rotational temperature of the three sources except for G10.30-0.15. The error originates from those of the rotational temperatures in the three sources.}
\end{deluxetable}

\section{Discussion}

\subsection{Comparisons of the rotational temperature of HC$_{5}$N} \label{sec:d1}

We compare the results in the high-mass star-forming regions with that in the cyanopolyyne peak in Taurus Molecular Cloud-1 (TMC-1 CP, $d=140$ pc), which is a well studied dark cloud and carbon-chain-species-rich source.
The derived excitation temperature of HC$_{5}$N in TMC-1 CP is $6.5 \pm 0.2$ K \citep{2016apj...817...147}.
The rotational temperatures of HC$_{5}$N in our three target high-mass star-forming regions except for G10.30-0.15, $\sim 13-20$ K, are significantly higher than the excitation temperature in the dark cloud TMC-1 CP.

The rotational temperature of HC$_{5}$N in the low-mass star-forming region L1527 ($d=140$ pc) is known to be $14.7 \pm 5.3$ K  \citep[$3\sigma$,][]{2009apj...702...1025}.
The rotational temperatures of HC$_{5}$N in our target high-mass star-forming regions are comparable with or slightly higher than that in L1527.
L1527 is one of the WCCC sources, in which carbon-chain species are formed in the warm gas around the protostar with the temperature of $\sim 20-30$ K \citep{2013ChRv...113...8981}.
Therefore, HC$_{5}$N seems to exist in the warm gas in our target high-mass star-forming regions.

The detection of the emission lines with the high excitation energies also allows us to confirm the warm components of HC$_{5}$N by itself.
The detected emission lines of HC$_{5}$N with the high-excitation energies ($E_{\rm {u}}/k \sim 90-100$ K) should not come from the cold environments with the gas kinetic temperature of 10 K.
We assume that the column density of HC$_{5}$N is comparable with that in TMC-1 CP \citep[$N$(HC$_{5}$N) = ($6.2 \pm 0.3$) $\times 10^{13}$ cm$^{-2}$,][]{2016apj...817...147} and the rotational temperature is a typical gas kinetic temperature in cold dark cloud (10 K).
The peak intensities of the $J= 38-37$ and 39$-$38 lines are calculated to be 0.8 mK in $T_{\rm {A}}^{*}$, even though the column density of the particular carbon-chain-rich source is considered.
Thus, the detection of the emission lines of HC$_{5}$N with the energies of $\sim 90-100$ K implies that HC$_{5}$N exists in the warm gas.
There is a possibility that the emission lines with the energies of $< 20$ K come from not only the warm gas but also the cold gas.
In that case, the derived rotational temperatures will be the lower limits due to the mixing of the warm gas and the cold gas in the single-dish beams.
To summarize, we can conclude that HC$_{5}$N exists in the warm gas within 0.07$-$0.1 pc radii around the massive young stellar objects.

\subsection{Possibility of chemical differentiation in the high-mass star-forming cores} \label{sec:diversity}

\subsubsection{$N$(HC$_{5}$N)/$W$(CH$_{3}$OH) ratios in the four high-mass star-forming regions} \label{sec:diversity1}

\begin{figure}
\figurenum{7}
\plotone{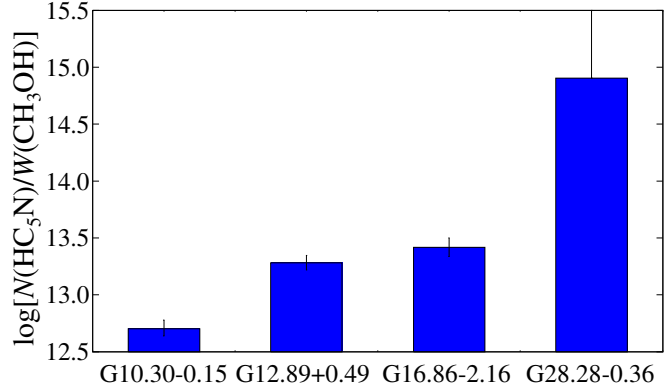}
\caption{Ratios of $N$(HC$_{5}$N)/$W$(CH$_{3}$OH) [(K km s$^{-1}$)$^{-1}$ cm$^{-2}$] in the four high-mass star-forming regions. The error bars show the standard deviation. The value of $W$(CH$_{3}$OH) in G28.28-0.36 is the $4 \sigma$ upper limit. \label{fig:f7}}
\end{figure}

We investigate the relationship between HC$_{5}$N and CH$_{3}$OH in the high-mass star-forming cores.
CH$_{3}$OH is one of the representative hot core tracers and COMs, and its origin is considered to be different from carbon-chain molecules.
Figure \ref{fig:f7} shows the comparison of the ratios between $N$(HC$_{5}$N) the column density of HC$_{5}$N and $W$(CH$_{3}$OH) the integrated intensity of the thermal CH$_{3}$OH emission line summarized in Table \ref{tab:tab2}.
We cannot derive the excitation temperature and column density of CH$_{3}$OH due to the detection of only single line with low-excitation energy ($E_{\rm {u}} < 50$ K).
The other detected lines with high-excitation energies in G12.89+0.49 could possibly be masers \citep{2004aap...428...1019}.
Therefore, we cannot use these lines to derive the excitation temperatures.

Since the column density of HC$_{5}$N in G10.30-0.15 was derived using its average rotational temperature of the other three high-mass star-forming regions, we do not discuss the result of G10.30-0.15 in detail due to its large uncertainty.
We can divide three high-mass star-forming regions into two groups; G28.28-0.36 shows the largest $N$(HC$_{5}$N)/$W$(CH$_{3}$OH) ratio of $> 8.0 \times 10^{14}$ in units of (K km s$^{-1}$)$^{-1}$ cm$^{-2}$, while those of G12.89+0.49 and G16.86-2.16 are only ($1.9 \pm 0.3$) $\times 10^{13}$ and ($2.6 \pm 0.5$) $\times 10^{13}$, respectively. 
The ratio in G28.28-0.36 is significantly larger than those in G12.89+0.49 and G16.86-2.16 by one order of magnitude, even though there is a possibility that the chemical and physical structures in the beams cause the observed large difference.
The $N$(HC$_{5}$N) values are almost comparable among the three sources.
Therefore, the differences in the $N$(HC$_{5}$N)/$W$(CH$_{3}$OH) ratio come from the $W$(CH$_{3}$OH) values.

The integrated intensity is proportional to both the column density and the rotational temperature.
When we assume the same $N$(HC$_{5}$N)/$N$(CH$_{3}$OH) ratios in all of the three sources, the differences in the $N$(HC$_{5}$N)/$W$(CH$_{3}$OH) ratios depend on only the rotational temperature of CH$_{3}$OH.
If the rotational temperature of CH$_{3}$OH in G28.28-0.36 is the typical value in hot cores \citep[100$-$250 K,][]{2007aap...465...913}, the large difference in the integrated intensities of CH$_{3}$OH by an order of magnitude (e.g., for the case of G28.28-0.36 and G16.86-2.16) cannot be explained given any rotational temperature assumed in G16.86-2.16. 
Alternatively, if the rotational temperatures of CH$_{3}$OH in G12.89+0.49 and G16.86-2.16 are 100$-$250 K, G28.28-0.36 will have an unlikely high rotational temperature of CH$_{3}$OH, 1200$-$2600 K.
Thus, the differences in the $N$(HC$_{5}$N)/$W$(CH$_{3}$OH) ratio among the three sources cannot be explained only by changes in the rotational temperature of CH$_{3}$OH, but the column density of CH$_{3}$OH in G28.28-0.36 should be lower than those in G12.89+0.49 and G16.86-2.16.
In summary, the observed variation in the $N$(HC$_{5}$N)/$W$(CH$_{3}$OH) ratio reflects the chemical differentiation in the single-dish beams.

\subsubsection{Fractional abundances of HC$_{5}$N in the four high-mass star-forming regions} \label{sec:diversity2}
 
We also derived the fractional abundances of HC$_{5}$N defined as $X$(HC$_{5}$N) = $N$(HC$_{5}$N)/$N$(H$_{2}$) toward the four sources.
We calculated the column densities of H$_{2}$ from the SCUBA, installed on the James Clerk Maxwell Telescope, 850 $\micron$ continuum data using the following formula \citep{2005...Kauffman};
\begin{equation} \label{H2}
N({\rm {H}}_2) = 2.02 \times 10^{20} {\rm {cm}}^{-2} \Bigl(e^{1.439(\lambda /{\rm {mm}})^{-1}(T/10 {\rm {K}})^{-1}}-1\Bigr)\Bigl(\frac{\kappa_{\nu}}{0.01\: {\rm {cm}}^{2}\: {\rm {g}}^{-1}}\Bigr)^{-1}\Bigl(\frac{S_{\nu}^{\rm {beam}}}{{\rm {mJy}}\: {\rm {beam}}^{-1}}\Bigr)\Bigl(\frac{\theta_{\rm {HPBW}}}{10\: {\rm {arcsec}}}\Bigr)^{-2}\Bigl(\frac{\lambda}{{\rm {mm}}}\Bigr)^{3}.
\end{equation}
The continuum data were obtained from the Canadian Astronomy Data Center, JCMT Science Archive\footnote{http://www.cadc-ccda.hia-iha.nrc-cnrc.gc.ca/en/jcmt/}.
We used the continuum fluxes ($S_{\nu}^{\rm {beam}}$) toward the observed positions and the beam size (18$^{\prime\prime}$) as summarized in Table \ref{tab:tab5}\footnote{Proposal IDs are m05ai02 for G10.30-0.15, m03bc05 for G12.89+0.49, and m00au01 for G16.86-2.16 and G28.28-0.36, respectively.}.
We used 0.0182 cm$^{2}$ g$^{-1}$ for $\kappa_{\nu}$ \citep{2005...Kauffman}, and 18$^{\prime\prime}$ for $\theta_{\rm {HPBW}}$.
$T$ is the dust temperature obtained by SED fitting \citep{2009MNRAS...394..323}, as summarized in Table \ref{tab:tab5}.
In G28.28-0.36, the dust temperature was not derived, and we used the average value of 58 K, determined from 65 good fits \citep{2009MNRAS...394..323}.
Table \ref{tab:tab5} summarizes the column densities of molecular hydrogen $N$(H$_{2}$) and the $X$(HC$_{5}$N) values.
The errors of $N$(H$_{2}$) are derived assuming that uncertainty of the dust temperatures is 20 K.

\begin{figure}
\figurenum{8}
\plotone{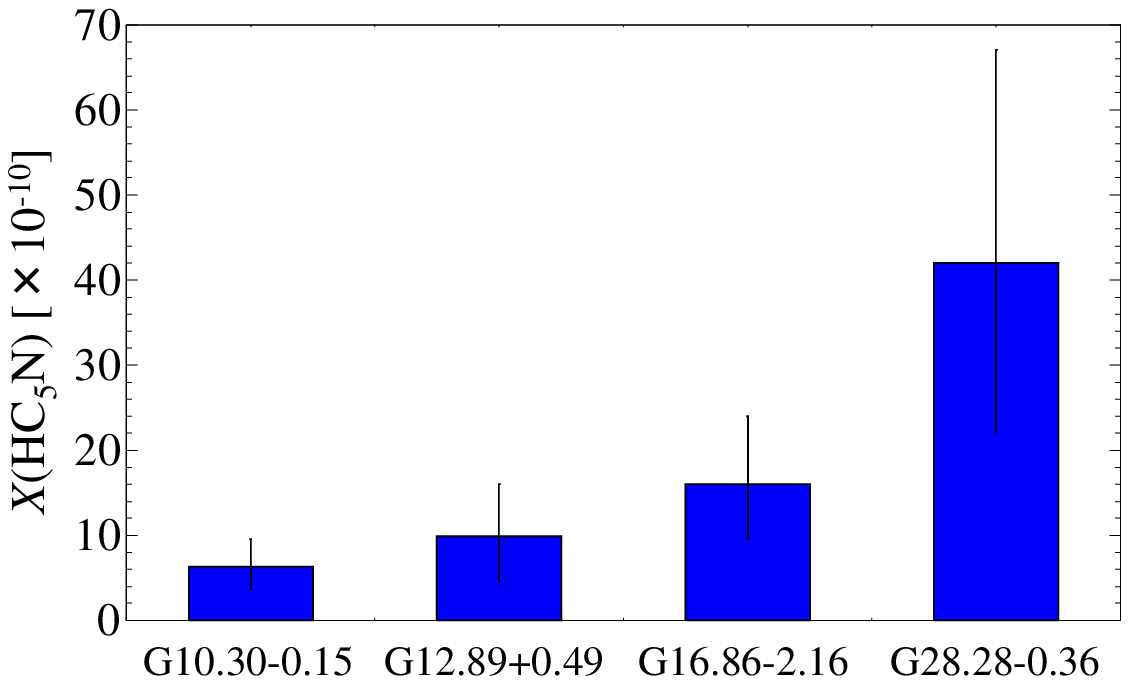}
\caption{Fractional abundances of HC$_{5}$N, $X$(HC$_{5}$N), in the four high-mass star-forming regions. The error bars show the standard deviation. \label{fig:f8}}
\end{figure}

Figure \ref{fig:f8} shows the comparison of the $X$(HC$_{5}$N) values among the four sources.
The $X$(HC$_{5}$N) values show the same tendency as the $N$(HC$_{5}$N)/$W$(CH$_{3}$OH) ratios.
G28.28-0.36 shows the larges $X$(HC$_{5}$N) value ($4.2 ^{+2.6}_{-2.0} \times 10^{-9}$) in the four sources, and G12.89+0.49 and G16.86-2.16 show the similar values, which is smaller than that in G28.28-0.36.
As mentioned before, the column density of HC$_{5}$N in G10.30-0.15 was derived using the average rotational temperature of the three sources, and so we do not discuss in detail.

Both Figures \ref{fig:f7} and  \ref{fig:f8} imply a possible signature of chemical differentiation among the high-mass star-forming cores, although we selected only four biased samples associated with HC$_{5}$N emission and the 6.7 GHz methanol masers. 
We cannot exclude a possibility of the chemical and physical structures in the beam, and one possible explanation for such chemical differentiation is the different beam filling factor of hot core regions \citep{1998A&AS..133...29H}.
Observations to reveal the spatial distributions of carbon-chain molecules and COMs with the interferometry are essential for revealing the origin of the chemical differentiation.

The fractional abundance of HC$_{5}$N in L1527 was derived to be $\sim 2 \times 10^{-10}$ ($N$(HC$_{5}$N) = ($6.8 \pm 1.4$)$\times 10^{12}$ cm$^{-2}$ and $N$(H$_{2}$) = $2.8 \times 10^{22}$ cm$^{-2}$ were derived by \citet{2009apj...702...1025} and \citet{2002A&A...389..908J}, respectively).
Even though there are uncertainties in column densities of H$_{2}$ and HC$_{5}$N due to assumed parameters (e.g. temperature, dust opacity, source size), the derive fractional abundances of HC$_{5}$N in the three high-mass star-forming regions (except for G10.30-0.15) are comparable to or slightly higher than that in L1527 by a factor of 5$-$20.
Such high HC$_{5}$N abundances in our target sources suggest that the warm gas around these massive young stellar objects is rich in HC$_{5}$N.
Moreover, the high HC$_{5}$N abundances would suggest efficient formation mechanisms of carbon chains in these high-mass star-forming regions (\citet{2008ApJ...674...993} and \citet{2008ApJ...681...1385} for WCCC and \citet{2009MNRAS...394...221} for hot core).

\floattable
\begin{deluxetable}{ccccc}
\tabletypesize{\scriptsize}
\tablecaption{The values of 850 $\micron$ continuum flux, dust temperature, $N$(H$_{2}$), and $X$(HC$_{5}$N) in the four hot cores\label{tab:tab5}}
\tablewidth{0pt}
\tablehead{
\colhead{} & \colhead{850 $\micron$ flux} & \colhead{$T$\tablenotemark{a}} & \colhead{$N$(H$_{2}$)\tablenotemark{b}} & \colhead{$X$(HC$_{5}$N)} \\
\colhead{Source} & \colhead{(Jy beam$^{-1}$)} & \colhead{(K)} & \colhead{($\times 10^{22}$ cm$^{-2}$)} & \colhead{($\times 10^{-10}$)} 
}
\startdata
G10.30-0.15 & 4.1 & 71 & $1.2 ^{+0.7}_{-0.3}$ & $6.3 ^{+3.2}_{-2.6}$ \\
G12.89+0.49 & 5.8 & 54 & $2.4 ^{+2.5}_{-0.8}$ & $9.9 ^{+6.1}_{-5.4}$ \\
G16.86-2.16 & 5.8 & 70 & $1.7 ^{+1.0}_{-0.5}$ & $16 \pm 7$ \\
G28.28-0.36 & 1.3 & 58 \tablenotemark{c} & $0.49 ^{+0.42}_{-0.16}$ & $ 42 ^{+26}_{-20} $ \\
\enddata
\tablecomments{The errors represent the standard deviation.}
\tablenotetext{a}{Taken from online material of \citet{2009MNRAS...394..323}.}
\tablenotetext{b}{The errors were derived from the uncertainties of the dust temperatures of 20 K.}
\tablenotetext{c}{The average value derived from 65 good fits \citep{2009MNRAS...394..323}.}
\end{deluxetable}

\section{Conclusions} \label{sec:con}

We carried out line survey observations at the 26$-$30 GHz band toward the four high-mass star-forming regions containing hot cores, G10.30-0.15, G12.89+0.49, G16.86-2.16, and G28.28-0.36, with the Robert C. Byrd Green Bank Telescope.
We have detected HC$_{5}$N from all of the sources, and HC$_{7}$N from the three sources except for G10.30-0.15.
The detection of HC$_{7}$N means that these high-mass star-forming regions contain plenty of long cyanopolyynes compared to typical star-forming regions.

We also conducted observations of HC$_{5}$N at the 42$-$46 GHz and 82$-$103 GHz bands toward the three high-mass star-forming regions, G12.89+0.49, G16.86-2.16, and G28.28-0.36, with the Nobeyama 45 m radio telescope.
We have detected the rotational lines of HC$_{5}$N with the high excitation energies ($E_{\rm {u}}/k \sim 63-100$ K), which are hardly excited in the cold dark clouds.
We conducted the rotational diagram analysis, combining the GBT data with the Nobeyama data.
The rotational temperatures of HC$_{5}$N in the three high-mass star-forming regions are derived to be $\sim 13-20$ K.
The detection of the lines with the high excitation energies and the derived rotational temperatures indicate that HC$_{5}$N exists in the warm gas within 0.07$-$0.1 pc radii around the massive young stellar objects.
The derived column densities of HC$_{5}$N in the three sources are ($\sim 2.0-2.8$) $\times 10^{13}$ cm$^{-2}$.

We compare the ratios between the column density of HC$_{5}$N, $N$(HC$_{5}$N), and the integrated intensity of the thermal CH$_{3}$OH emission line, $W$(CH$_{3}$OH) among the three high-mass star-forming regions.
We found a possibility of the chemical differentiation in the three high-mass star-forming regions; G28.28-0.36 shows the largest $N$(HC$_{5}$N)/$W$(CH$_{3}$OH) ratio of $> 8.0 \times 10^{14}$ in units of (K km s$^{-1}$)$^{-1}$ cm$^{-2}$, while G12.89+0.49 and G16.86-2.16 show the smaller values ($\sim 2 \times 10^{13}$). 
Such chemical differentiation may originate from the complex chemical and physical structures in the beam.

\acknowledgments
We deeply appreciate the anonymous referee for constructive suggestions, which are very useful to develop our discussion.
We would like to express our great thanks to the staff members of the National Radio Astronomy Observatory.
The National Radio Astronomy Observatory is a facility of the National Science Foundation operated under cooperative agreement by Associated Universities, Inc.
K. T. deeply appreciates Dr. Nichol Cunningham (NRAO) for her kind help to prepare and carry out observations. 
We would like to express our thanks to the staff members of the Nobeyama Radio Observatory.
The Nobeyama Radio Observatory is a branch of the National Astronomical Observatory of Japan, National Institutes of Natural Sciences.
%

\vspace{5mm}
\facilities{Robert C. Byrd Green Bank Telescope, Nobeyama 45 m radio telescope}


\software{GBTIDL, Java Newstar}


\begin{thebibliography}{}
\bibitem[Aikawa et al.(2008)]{2008ApJ...674...993} Aikawa, Y., Wakelman, V., Garrod, R. T., \& Herbst, E.\ 2008, \apj, 674, 993
\bibitem[Alexander et al.(1976)]{1976JMoSp...62...175} Alexander, A. J., Kroto, H. W., \& Walton, D. R. M.\  1976, JMoSp, 62, 175
\bibitem[Bergin et al.(1996)]{1996ApJ...460...343} Bergin, E. A., Snell, R. L., \& Goldsmith, P. F.\ 1996, \apj, 460, 343
\bibitem[Bisschop et al.(2007)]{2007aap...465...913} Bisschop, S. E., J$\o$rgensen, J. K., van Dishoeck, E. F., \& de Wachter, E. B. M.\  2007, \aap, 465, 913
\bibitem[Chapman et al.(2009)]{2009MNRAS...394...221} Chapman, J. F., Millar, T. J., Wardle, M., Burton, M. G., \& Walsh, A. J.\ 2009, \mnras, 394, 221
\bibitem[Cordiner et al.(2012)]{2012ApJ...744..131C} Cordiner, M. A., Charnley, S. B., Wirstr{\"o}m, E. S., \& Smith, R. G.\ 2012, \apj, 744, 131 
\bibitem[Cyganowski et al.(2008)]{2008AJ...136...2391} Cyganowski, C. J., Whitney, B. A., Holden, E., et al.\  2008, \aj, 136, 2391
\bibitem[Feng et al.(2015)]{2015aap...581...A71} Feng, S., Beuther, H., Henning, Th., et al.\  2015, \aap, 581, A71
\bibitem[Friesen et al.(2013)]{2013MNRAS.436.1513F} Friesen, R. K., Medeiros, L., Schnee, S., et al.\ 2013, \mnras, 436, 1513
\bibitem[Garrod \& Herbst(2006)]{2006A&A...457...927} Garrod, R. T., \& Herbst, E.\ 2006, \aap, 457, 927
\bibitem[Goddi et al.(2011)]{2011apj...739...L13} Goddi, C., Greenhill, L. J., Humphreys, E. M. L., Chandler, C. J., \& Matthews, L. D.\ 2011, \apjl, 739, L13
\bibitem[Goldsmith \& Langer(1999)]{1999ApJ...517..209G} Goldsmith, P.~F., \& Langer, W.~D.\ 1999, \apj, 517, 209
\bibitem[Green et al.(2014)]{2014MNRAS...443...2252} Green, C. -E., Green, J. A., Burton, M. G., et al.\  2014, \mnras, 443, 2252
\bibitem[Hassel et al.(2008)]{2008ApJ...681...1385} Hassel, G. E., Herbst, E., \& Garrod, R. T.\ 2008, \apj, 681, 1385
\bibitem[Hatchell et al.(1998)]{1998A&AS..133...29H} Hatchell, J., Thompson, M. A., Millar, T. J., \& MacDonald, G. H.\ 1998, \aaps, 133, 29
\bibitem[Herbst \& van Dishoeck(2009)]{2009araa...47...427} Herbst, E., \& van Dishoeck, E. F.\  2009, \araa, 47, 427
\bibitem[Hirota et al.(2009)]{2009apj...699...585} Hirota, T., M. Ohishi, \& Yamamoto, S\  2009, \apj, 699, 585
\bibitem[J{\o}rgensen et al.(2002)]{2002A&A...389..908J} J{\o}rgensen, J. K., Sch{\"o}ier, F. L., \& van Dishoeck, E. F.\ 2002, \aap, 389, 908
\bibitem[Kamazaki et al.(2012)]{2012PASJ...64...29K} Kamazaki, T., Okumura, S.~K., Chikada, Y., et al.\ 2012, \pasj, 64, 29
\bibitem[Kauffmann(2005)]{2005...Kauffman} Kauffmann, J.\  2005, Column Densities and Masses from Dust Emission, http://youngstars.nbi.dk/jeskj/AstroSpectra/kauffmann05.pdf
\bibitem[Li et al.(2016)]{2016aj...152...92L} Li, F. C., Xu, Y., Wu, Y. W., et al.\  2016, \aj, 152, 92
\bibitem[M$\ddot{\rm u}$ller et al.(2004)]{2004aap...428...1019} M$\ddot{\rm u}$ller, H. S. P., Menten, K. M., \& M$\ddot{\rm a}$der, H.\ 2004, \aap, 428, 1019
\bibitem[M$\ddot{\rm u}$ller et al.(2005)]{2005JMoSt...742...215} M$\ddot{\rm u}$ller, H. S. P., Schl$\ddot{\rm o}$der, F., Stutzki, J., \& Winnewisser, G.\  2005, JMoSt, 742, 215
\bibitem[Nakajima et al.(2013)]{2013PASP..125..252N} Nakajima, T., Kimura, K., Nishimura, A., et al.\ 2013, \pasp, 125, 252
\bibitem[Nakamura et al.(2015)]{2015PASJ...67..117N} Nakamura, F., Ogawa, H., Yonekura, Y., et al.\ 2015, \pasj, 67, 117
\bibitem[Purcell et al.(2006)]{2006MNRAS...367...553} Purcell, C. R., Balasubramanyam, R., Burton, M. G., et al.\  2006, \mnras, 367, 553
\bibitem[Purcell et al.(2009)]{2009MNRAS...394..323} Purcell, C. R., Longmore, S. N., Burton, M. G., et al.\ 2009, \mnras, 394, 323
\bibitem[Reid et al.(2014)]{2014apj...783...130} Reid, M. J., Menten, K. M., Brunthaler, A., et al.\  2014, \apj, 783, 130
\bibitem[Sakai et al.(2008)]{2008apj...672...371} Sakai, N., Sakai, T., Hirota, T., \& Yamamoto, S.\  2008, \apj, 672, 371
\bibitem[Sakai et al.(2009a)]{2009apj...697...769} Sakai, N., Sakai, T., Hirota, T., Burton, M., \& Yamamoto, S.\  2009a, \apj, 697, 769
\bibitem[Sakai et al.(2009b)]{2009apj...702...1025} Sakai, N., Sakai, T., Hirota, T., \& Yamamoto, S.\ 2009b, \apj, 702, 1025
\bibitem[Sakai et al.(2010)]{2010ApJ...722.1633} Sakai, N., Sakai, T., Hirota, T., \& Yamamoto, S.\ 2010, \apj, 722, 1633
\bibitem[Sakai \& Yamamoto(2013)]{2013ChRv...113...8981} Sakai, N., \& Yamamoto, S.\  2013, ChRv, 113, 8981
\bibitem[Suzuki et al.(1992)]{1992apj...392...551} Suzuki, H., Yamamoto, S., Ohishi, M., et al.\  1992, \apj, 392, 551
\bibitem[Taniguchi et al.(2016a)]{2016apj...817...147} Taniguchi, K., Ozeki, H., Saito, M., et al.\  2016a, \apj, 817, 147
\bibitem[Taniguchi et al.(2016b)]{2016apj...830...106} Taniguchi, K., Saito, M., \& Ozeki, H.\  2016b, \apj, 830, 106
\bibitem[Thompson et al.(2006)]{2006aap...453...1003} Thompson, M. A., Hatchell, J., Walsh, A. J., Macdonald, G. H., \& Millar, T. J.\  2006, \aap, 453, 1003
\bibitem[Walsh et al.(2003)]{2003aap...410...597} Walsh, A. J., Macdonald, G. H., Alvey, N. D. S., Burton, M. G., \& Lee, J.\  2003, \aap, 410, 597 
\end{thebibliography}
\end{document}